\documentclass[preprint,preprintnumbers,amsmath,amssymb]{revtex4}
\usepackage{amsfonts}
\usepackage{amssymb}
\usepackage{latexsym}
\newcommand{\al}{\alpha}

\newcommand{\Si}{\Sigma}

\newcommand{\De}{\Delta}

\newcommand{\lrar}{\leftrightarrow}

\begin{document}


\title{A note about the ground state of the hydrogen molecule}

\author{A.V.~Turbiner}
\email{turbiner@nucleares.unam.mx}

\author{N.L.~Guevara}
\email{nicolais@nucleares.unam.mx}
\affiliation{Instituto de Ciencias Nucleares, Universidad Nacional
Aut\'onoma de M\'exico, Apartado Postal 70-543, 04510 M\'exico,
D.F., Mexico}


\begin{abstract}
A trial function is presented for the $H_2$ molecule which provides
the most accurate (the lowest) Bohr-Oppenheimer ground state energy
among few-parametric trial functions (with $\leq 14$ parameters). It
includes the electronic correlation term in the form $\sim
\exp{(\gamma r_{12})}$ where $\gamma$ is a variational parameter.
\end{abstract}
%

\maketitle

\section{Introduction}

Hydrogen molecule $H_2$ is among the most important chemical objects
which appear in Nature. Since early days of quantum mechanics after
pioneering paper by James and Coolidge \cite{JC:1933} many studies
of $H_2$ were carried out (see \cite{Sims-Hangstrom:2006} and
references therein). The paper \cite{JC:1933} contained a clear
indication that the interelectron correlation must be included
explicitly. In general, the success of calculations and, in
particular, a rate of convergence of a method used depends very much
on a form of the correlation factor
\cite{Bressanini,Ten-no:2004,Manby:2006} (for a review, see
\cite{Klopper:2006}, Section 2.2). In particular, recently, it was
drawn a conclusion based on an analysis of many atomic and molecular
systems that the best form of correlation factor is $\exp(\gamma
r_{12})$ comparing to linear or the Gaussian in $r_{12}$ factors
\footnote{It is worth noting that any method based on the Gaussian
in $r_{12}$ factor (or a superposition of the Gaussian functions as
a factor) has an evident drawback - by construction the cusp
parameter $\frac{d \log \Psi}{dr_{12}}\vert_{r_{12}=0}$ is {\it
always} zero (see, e.g., \cite{Bressanini}). The fact, that those
methods can lead to very accurate results for the energy, implies a
relative {\it unimportance} of the short-range behavior in $r_{12}$
of the trial functions for getting the high-precision results in the
calculation of the BO ground state energy. It also implies that
these methods are doomed to a low accuracy of the expectation values
for which a short-range behavior in $r_{12}$ of the wavefunction is
important. This situation can be easily explored taking the H-atom
as an example and applying a superposition of the Gaussian functions
as a trial function \cite{H}}. No clear reason was given so far why
it is so. A goal of this note is to present a simple, compact,
easy-to-handle trial function which leads to the most accurate (the
lowest) Bohr-Oppenheimer ground state energy among few-parametric
trial functions ($\leq 14$ parameters). The variational energy is
calculated numerically using a specially designed computer code for
multidimensional numerical integration with high accuracy. The trial
function contains interelectron correlation in the form $\exp(\gamma
r_{12})$. It is worth mentioning that long time ago this dependence
on $r_{12}$ appeared in the variational trial functions in studies
of the $H_2$ molecule in a magnetic field \cite{Turbiner:1983} and,
recently, of other two-electron molecular systems in a magnetic
field \cite{Turbiner:2006_h3,Turbiner:2006_he2ee,Turbiner:2006}. A
hint why namely this $r_{12}$-dependence leads to the fast
convergent results will be given.

The Hamiltonian which describes the hydrogen molecule under the
assumption that the protons are infinitely massive (the
Born-Oppenheimer approximation of zero order) can be written as
follows
\begin{equation}
\label{H}
  {\cal H}\ =\sum_{\ell=1}^2 {\hat {\mathbf p}_{\ell}}^2\ -
  \ \sum_{\buildrel{{\ell}=1,2}\over{\kappa =A,B}}
  \frac{2}{r_{{\ell},\kappa}}
  \ +\ \frac{2}{r_{12}}\ +\ \frac{2}{R}  \ ,
\end{equation}
where ${\hat {\mathbf p}_{\ell}}=-i \nabla_{\ell}$ is the 3-vector
of the momentum of the ${\ell}$th electron, the index $\kappa$ runs
over protons $A$ and $B$, $r_{{\ell},\kappa}$ is the distance
between ${\ell}$th electron and $\kappa$ proton, $r_{12}$ is the
interelectron distance, $R$ is the interproton distance.
It is the established fact that the ground state of the $H_2$
molecule is ${}^1 \Si_g^+$, the spin-singlet state, symmetric under
permutations of electron positions as well as proton positions.

\section{Variational method}

The variational procedure is used as a method to explore the
problem. The recipe of choosing the trial function is based on a
physical relevance arguments (see, e.g. \cite{turbinervar}). In
practice, use of such trial functions implies the convergence of a
special form of the perturbation theory where the variational energy
is the sum of the first two terms. Let us remind the essentials of
this perturbation theory (for details, see \cite{turbinervar}). Let
us assume that our original Hamiltonian has a form ${\cal H}=-\Delta
+ V$. As a first step we choose a trial function $\psi^{(trial)}$
which is normalized to one. Then we find a potential for which our
trial function $\psi^{(trial)}$ is the exact eigenfunction
$V_{trial}=\frac{\De \psi^{(trial)}}{\psi^{(trial)}}$ with the
energy $E_{trial}=0$. In a pure formal way we can construct a
Hamiltonian ${\cal H}_{trial} = -\Delta + V_{trial}$ such that
${\cal H}_{trial} \psi^{(trial)}=0$. It can be easily shown that the
variational energy
\[
 E_{var}= <\psi^{(trial)}|{\cal H}|\psi^{(trial)}>
\]
is nothing but the first two terms in the perturbation theory where
the unperturbed problem is given by ${\cal H}_{trial}$ and the
perturbation is the deviation of the original potential $V$ from the
trial potential $V_{trial}$, namely, $V_{perturbation}=V-V_{trial}$.
Eventually, we arrive at the formula
\begin{equation}
    E_{var}= E_{trial} + E_1 (V_{perturbation})\ ,
\end{equation}
here $E_1
(V_{perturbation})=<\psi^{(trial)}|V_{perturbation}|\psi^{(trial)}>$
is the first energy correction in the perturbation theory, where
unperturbed potential is $V_{trial}$. It is worth noting that if the
trial function is the Hartree-Fock function the resulting
perturbation theory is nothing but the Moeller-Plesset perturbation
theory (see, e.g. \cite{Levine}, Section 15.18) \footnote{It is
worth noting that the question about a convergence of the Moeller
Plesset perturbation theory is not settled yet \cite{MP}}.

One of the criteria of convergence of the perturbation theory in
$V_{perturbation}=V-V_{trial}$ is a requirement that the ratio
$|\frac{V_{perturbation}}{V}|$ should not grow when $r$ tends to
infinity in any direction. If this ratio is bounded by a constant it
should be less than one. In fact, it is a condition that the
perturbation potential is subordinate with respect to the
unperturbed potential. A value of this constant controls the rate of
convergence - a smaller value of this constant leads to faster
convergence \cite{turbinervar1}. Hence, the above condition gives a
importance to the large-range behavior of the trial functions. In
the physics language the above requirement means that the phenomenon
of the Dyson's instability should not occur (for a discussion see
\cite{turbinervar}) \footnote{It is worth noting that this procedure
for a selection of the trial function was applied successfully to a
study of one-electron molecular systems in a magnetic field leading
to the highly accurate results. Many of these results are the most
accurate at the moment (see \cite{turbiner:2006}).}. Among three
factors which are mentioned in literature (see \cite{Klopper:2006}):
the linear in $r_{12}$, exponential $\exp(\gamma r_{12})$ and
$\exp(-\alpha r_{12}^2)$, the only factor $\exp(\gamma r_{12})$
fulfills the above requirement. It was demonstrated in
\cite{Bressanini} that a superposition of the Coulomb functions with
exponentially correlated function $\exp(\gamma r_{12})$ (see below,
eq.(4)) leads to faster convergence than others. Perhaps, it is
worth mentioning that for the case of Gaussian factor the
above-defined constant is equal to one exactly. In concrete, by
following the above procedure and a requirement of the convergence
of the perturbation theory we choose the trial function for the
ground state in a form
\begin{equation}
\label{ansatz}
 \psi^{(trial)} =A_1 \psi_1 + A_2 \psi_2 + A_3 \psi_3
 \end{equation}
where
\begin{equation}
\label{ansatz1}
 \psi_1 = (1+ P_{12})(1+ P_{AB}) e^{
 -\al_1 r_{1A}-\al_2 r_{1B} -\al_3 r_{2A} -\al_4 r_{2B}
 + \gamma_1 r_{12} }\, ,
 \end{equation}
 \begin{equation}
\label{ansatz2}
 \psi_2 = (1+ P_{12}) e^{ -\al_5 (r_{1A}+ r_{2B})-\al_6 (r_{1B}+r_{2A})
 + \gamma_2 r_{12} }\, ,
 \end{equation}
 \begin{equation}
\label{ansatz3}
 \psi_3 = (1+ P_{12}) e^{ -\al_7 (r_{1A}+ r_{1B})-\al_8 (r_{2A}+r_{2B})
 + \gamma_3 r_{12} }\, ,
\end{equation}
The $P_{12}$ is the operator which interchanges electrons $(1 \lrar
2)$ and $P_{AB}$ is the operator which interchanges the two nuclei
$A \lrar B$. It is easy to check that the functions (5)-(6) are
symmetric with respect to the interchange $A \lrar B$. The
variational parameters consist of non-linear parameters $\al_{1-8}$,
$\gamma_{1-3}$ which characterize (anti)screening of the Coulomb
charges and linear parameters $A_{1-3}$. If the internuclear
distance $R$ is taken into account the trial function (\ref{ansatz})
depends on 14 parameters \footnote{Due to a freedom in
  normalization of the wave function one of the coefficients $A$
  can be kept fixed; thus, in present calculation we put $A_1=1$}.
It is worth mentioning that (\ref{ansatz2}) is a degeneration of
(\ref{ansatz1}) when $\al_1=\al_4, \al_2=\al_3$ and (\ref{ansatz3})
is another degeneration of (\ref{ansatz1}) when $\al_1=\al_2,
\al_3=\al_4$. In a certain way, the function (\ref{ansatz2}) mimics
the interaction of two hydrogen atoms $H + H$, while the function
(\ref{ansatz3}) mimics the interaction $H_2^+ + e$. Eventually, the
function (\ref{ansatz1}) can be treated as a non-linear
interpolation between (\ref{ansatz2}) and (\ref{ansatz3}). Those
functions look analogous to the Hund-Mulliken, Heitler-London and
Guillemin-Zener functions, respectively.

Calculations were performed using the minimization package MINUIT
from CERN-LIB. Multidimensional integration was carried out
numerically using a "state-of-the-art" dynamical partitioning
procedure: a domain of integration was divided into subdomains
following an integrand profile, in particular, the domains with
sharp changes of the integrand were separated out. Then each
subdomain was integrated separately with controlled accuracy (for
details, see e.g. \cite{turbiner:2006}). A realization of the
routine requires a lot of attention and care. During minimization
process a partitioning was permanently controlled and adjusted.
Numerical integration was done with a relative accuracy of $\sim
10^{-6} - 10^{-7}$ by use of the adaptive D01FCF routine from
NAG-LIB. Computations were performed on a dual DELL PC with two Xeon
processors of 2.8\,GHz each.

\section{Results}

Present results for the ground state of the $H_2$ molecule and their
comparison with results of previous studies are displayed in Table
\ref{table1}. The Bohr-Oppenheimer ground state energy obtained
using the function (3)-(6) is the most accurate (the lowest) among
those obtained with other trial functions with $\leq 14$ parameters.
A reasonable agreement for expectation values is also observed,
except for $< 3z_1^2-r_1^2>$ related to the quadrupole moment. To
present authors it seems evident that this expectation value should
be studied separately (see also \cite{kolos}). It is not surprising
that the obtained value of the cusp parameter in $r_{12}$ is equal
to 0.4 unlike to the exact value 0.5 (see footnote [19]).
Variational parameters of the trial function (\ref{ansatz}) are
shown in Table~\ref{table2}. It is worth emphasizing that a
numerical calculations are very difficult and can easily lead to a
loss of accuracy. In \cite{Bressanini} a similar function (3) but
with all three components of the form (4) containing 18 variational
parameters was studied using variational Monte-Carlo technique. A
comparison of our results with less parameters with
\cite{Bressanini} (see Table I) indicates that we obtain a lower
total energy of the order $5 \times 10^{-4}$\,Ry.

\section{Conclusion}

We presented a simple and compact few-parametric trial function
which provides the most accurate Bohr-Oppenheimer energy for $H_2$
molecule among those based on few-parametric $(\leq 14)$ trial
functions. Emerging five-dimensional integrals were effectively
calculated using fast state-of-the-art integration routine which
admits parallelization. The trial function (\ref{ansatz}) can be
easily generalized by adding other physically-natural degenerations
of (\ref{ansatz1}) than (\ref{ansatz2}),(\ref{ansatz3}). One of them
appears when all $\al$-parameters in (\ref{ansatz1}) are equal. It
should be dominant in a domain of small interproton distances. It
seems natural to assume that taking linear superpositions of the
functions (\ref{ansatz1}) we end up with fast convergent procedure
(see \cite{Bressanini}). The function (\ref{ansatz}) can be easily
as modified for a study of spin-triplet states and as well as the
states of the lowest energy with non-vanishing magnetic quantum
numbers. A generalization to more-than-two electron systems is
straightforward.

\section*{Acknowledgement}

This note is dedicated to the memory of Professor Koutecky with whom
the first author had a privilege to talk extensively on several
occasions in his very last years. It was very striking to see his
openness and readiness to discuss new approaches in quantum
chemistry. In fact, it was his encouragement that finally led to the
present study.

The authors express their sincere gratitude to J.C.~L\'opez Vieyra
for the interest to the work and numerous useful discussions. The
arguments and support by J.~Cizek which convinced the present
authors to publish presented results are highly appreciated.

This work was supported in part by PAPIIT grant {\bf IN121106}
(Mexico). The first author is grateful to the University Program
FENOMEC (UNAM, Mexico) for a support.

\begin{table}
  \centering
\caption{\label{table1} Total energy $E_T$ in Ry and expectation
                        values in a.u. of the hydrogen molecule $H_2$
                        for the ground state. $r_1, z_1$ are
                        distances from 1st electron to the mid-point
                        between protons. Some data are rounded.}
\begin{tabular}{|c|c|c|c|c|c|}
\hline
   $E_T$ (Ry)  & $<r_{12}^{-1}>$ & $<r_{1}^2>$ & $<\frac{(r_{1A}+r_{1B})}{R}>$
   & $< 3z_1^2-r_1^2>$ &  Refs.\\
\hline
 -2.34697   \footnote{from \cite{JC:1933}
   (the BO energy with 14 variational parameters)}  &
          &         &        &        & \cite{JC:1933} \\
 -2.34778  \footnote{from \cite{Schmelcher} (based on use of $> 200$ non-spherical
   Gaussian orbitals)} &
          &&&& \cite{Schmelcher} \\
-2.34787  \footnote{from \cite{Bressanini} ($N=3$ exponentially
correlated functions)} &
          &&&& \cite{Bressanini} \\
 -2.348382   \footnote{from Table III \cite{kolos}
   (the BO energy with 14 variational parameters)} &
          & 2.5347  &        & 0.5227 & \cite{kolos} \\
 -2.348393 \footnote{Present calculation (14 variational parameters)} &
           0.5874  & 2.5487  & 2.2133 & 0.4847 & present \\
 -2.34872   \footnote{from Table III \cite{kolos}
   (the BO energy with 28 variational parameters )}  &
          & 2.5426  &        & 0.5142 & \cite{kolos}\\
 -2.34888 \footnote{from Table II \cite{kolos}
   (the BO ground state energy with 40 variational parameters)}
          & 0.5874  &         & 2.2127 &  & \cite{kolos} \\
 -2.34895   \footnote{from \cite{Sims-Hangstrom:2006}
   (7034 James-Coolidge type terms, the record calculation at present,
                   the number in Table is rounded))} &
          &         &        &        & \cite{Sims-Hangstrom:2006}\\
\hline
\end{tabular}
\end{table}

\begin{table}
  \centering
\caption{\label{table2} Parameters of the trial function
                        (\ref{ansatz}). }
\begin{tabular}{|c|c|}
\hline
\tableline
    R         &  1.40053    \\
\hline
    $A_1$     &  1.            \\
   $\alpha_1$    &  0.720674986   \\
   $\alpha_2 $   &  0.521577488   \\
   $\alpha_3 $   &  0.130799743   \\
   $\alpha_4 $   &  1.30816746    \\
   $\gamma_1 $   &  0.0655006315  \\
\hline
   $A_2$         & -1.15105579    \\
   $\alpha_5 $   &  0.604583808   \\
   $\alpha_6 $   &   0.658402827   \\
   $\gamma_2 $   & -0.349101361   \\
\hline
   $A_3$         &  0.256342676   \\
   $\alpha_7 $   &  0.968330781   \\
   $\alpha_8 $   &  0.229153253   \\
   $\gamma_3 $   & -0.354509413   \\
\tableline\hline
\end{tabular}
\\
\end{table}

\end{document}